\let\nopictures=Y
\documentstyle[12pt]{article}
\begin{document}
\begin{flushright}
hep-ph/9502309\\
FTUAM 94/30  \\
FTUCM 94/20 \\
CMS TN /94-276\\
February 1995
\end{flushright}
\vskip 1cm
\begin{center}
{\Large \bf Learning about the strongly interacting symmetry breaking
sector at LHC}	\\
\vskip 1cm
A. Dobado$^1$, M. J. Herrero$^2$, J. R. Pel\'aez$^1$, E. Ruiz
Morales$^2$ and  M.T. Urdiales$^2$
\vskip 0.1cm
{\small $^1$ Departamento de F\'{\i}sica Te\'orica, Universidad
Complutense, 28040 Madrid, Spain\\   $^2$ Departamento de F\'{\i}sica
Te\'orica, Universidad Aut\'onoma, 28049 Madrid, Spain }
\end{center}
\vskip 3cm
\begin{abstract}
In the present work we study the predictions for $WZ$ and $ZZ$
production at LHC with the Electroweak Chiral Lagrangian (EChL)
approach. Our analysis will be focused on the less favored case from
the experimental point of view, in which the predictions for the gauge
bosons scattering amplitudes are considered in the low energy range
where, by construction of the low energy approach, they reveal no
resonant
behavior. The study includes the complete set of amplitudes for all the
polarization states of the initial and/or final gauge bosons and
makes no use of the Equivalence Theorem. We express the results in
terms of the range of values of the chiral parameters that will be
accessible at LHC.
\end{abstract}
\newpage
\section{Introduction}
One of the main goals of the LHC is to get as much information as
possible about the electroweak symmetry breaking sector (ESBS) of the
Standard Model (SM).  If this sector is weakly interacting, some light
modes below the TeV energy regime are expected to appear. The typical
examples are the SM with a light Higgs particle and the Minimal
Supersymmetric Standard Model. In contrast, a strongly interacting
scenario is characterized by the absence of light modes.  In this
case, general considerations lead to the identities called Low Energy
Theorems (LET) \cite{1} that allow us to predict the general behavior
of the Goldstone boson amplitudes irrespective of the details of the
symmetry breaking mechanism. In fact, a very powerful theoretical
framework has been developed in the last years, which provides a
systematic phenomenological description of the ESBS in the strongly
interacting case.  This description \cite{2} is inspired in Chiral
Perturbation Theory (ChPT) which is known to work very well
in low energy pion
physics \cite{3}. The Chiral Lagrangian that is used to study the
ESBS is $SU(2)_L\times U(1)_Y$ gauge invariant, $CP$ conserving  and
includes effective operators up to dimension four \cite{4}.  It will
be referred here as the Electroweak Chiral Lagrangian (EChL).  This
approach incorporates from the beginning and by construction the LET
in a model independent way.  The details about the underlying ESBS
physics are encoded in the values of the couplings or parameters
$\alpha_i$ of the EChL which, hopefully, will be measured at LHC.  By
choosing properly these parameters one can reproduce different
strongly interacting scenarios, as for instance the SM case with a
heavy Higgs boson \cite{5,6}, Technicolor models \cite{7}, the BESS
model \cite{8}, etc.  The typical values of the chiral parameters in
most of these scenarios are $\alpha_i\leq 10^{-2}$.

In the previous applications of the EChL to the LHC physics
\cite{9,10} the Equivalence Theorem (ET) played an essential role.
This theorem \cite{11} relates the $S$ matrix elements of the
longitudinal components of the gauge bosons with the corresponding
ESBS Goldstone bosons, at energies much higher than $M_W$, thus
simplifying enormously the practical computations. However, the fact
that the ET should be applied only at high energies and the
intrinsic low-energy character of the EChL put severe limits on
their simultaneous application \cite{12,13}.  To solve this problem
some other extra non-perturbative methods are needed such as
unitarization procedures \cite{14}, dispersion relations  \cite{15}
or the large $N$ limit \cite{16}. With any of these methods  the use
of the ET in the EChL approach is possible and, in addition, they can
provide a description of resonant behavior in different channels.
This kind of approach was followed in \cite{9}, where the use of the
EChL, together with the ET and the Pad\'e unitarization method
allowed to describe two typical strongly interacting scenarios.
Incidentally, the use of the Pad\'e approximants seems to  be the
most reliable method as it has been tested in low energy pion physics
and, in addition, it can be rigorously justified from dispersion
relations \cite{15}. The two mentioned scenarios are the SM with a
Higgs-like scalar resonance that reveals mainly in $Z_LZ_L$
production and the so-called  QCD-like scenarios with a vector
resonance (it includes the case of the technirho resonance) that
emerges clearly in the $W^{\pm}_LZ_L$ channel.	For more details on
this resonant model we refer the reader to \cite{8}.

All the resonant models studied so far, however, have the
disadvantage of giving predictions only for longitudinal gauge bosons
and of being valid just for energies much higher than $M_W$, in order
to make
the ET be a reliable approximation. This, in practice, implies a
quite restrictive cut in the lowest invariant mass of the $VV$ gauge
boson pair of about $500 GeV$ and, in consequence, a significant lose
in the signal rates.

We will follow here a different approach to study the ESBS at LHC
which has been proposed in \cite{17} and we refer the reader to this
work for a full description of the method and for a more detailed
analysis. The method {\it makes no use of the ET} nor any
unitarization prescription but uses the EChL directly to compute the
amplitudes for all the polarization states of the initial and/or
final gauge bosons. It is technically more involved but is more
complete than previous studies where just longitudinal polarizations
were considered.  Furthermore, it has the advantage that there are no
restrictions on the lower end of the $VV$ invariant mass since the ET
is not used.  The only limitation is that it must be applied in the
low energy region, namely well below $4\pi v \sim 3\;TeV$
($v=246GeV$), in order for the low energy effective theory defined by
the
EChL to give reliable predictions. On the other hand,
for the
energies considered here which in practice will be imposed by the
kinematical cuts to lay below 1.5 TeV, the
predictions for the gauge bosons scattering amplitudes
reveal no resonant behavior. This
situation with no resonances showing up
is, in principle, more difficult to be tested experimentally.

The aim of this paper is, in summary, to study the possibilities for
meassuring the EChL parameters at LHC within this non-resonant EChL
model.	In particular we have chosen to study pair production of
$W^{\pm}Z$ and $ZZ$ gauge bosons with the $W's$ and $Z's$ decaying
into the cleanest leptonic ('gold-plated') channels: $W\rightarrow
\nu_ee,\nu_{\mu}\mu$ and $Z\rightarrow e^+e^-,\mu^+\mu^-$. The
results will be expressed in terms of the range of values of the
chiral parameters that will be accessible at LHC by means of these
two channels.

\section{The electroweak chiral parameters}

Let us start fixing the notation for the electroweak chiral
parameters, $\alpha_i$, which are the object of our study. These
parameters appear in the definition of the EChL
which is made of the complete set of $SU(2)
\times U(1)_Y$, Lorentz, $C$, and $P$ invariant operators up to
dimension four. The EChL is given by
\begin{equation}
{\cal L}_{EChL}= {\cal L}_{NL}+ \sum_{i=0}^{13} {\cal L}_i
\label{EChL}
\end{equation}
where ${\cal L}_{NL}$ is the Lagrangian of the
gauged non-linear sigma model
\begin{equation}
{\cal L}_{NL}= \frac{v^2}{4}tr[D_{\mu} U(D^{\mu}U)^{\dagger}]
	-\frac{1}{4} B_{\mu\nu} B^{\mu\nu} - \frac{1}{2}
	Tr[F_{\mu\nu}F^{\mu\nu}]+{\cal L}_{GF}+{\cal L}_{FP}
\label{NL}
\end{equation}
which is written down in terms of a non-linear parametrization of the
would-be Goldstone boson (GB) fields $\pi_i$, and the electroweak
gauge boson fields ${\vec{W}}_{\mu}$ and $B_{\mu}$
\begin{equation}
U = {\rm exp} (i \frac{\vec{\tau}.\vec{\pi}}{v})
\;\;,\;\;
W_{\mu} = \frac{ {\vec{W}}_{\mu} {\vec{\tau}}}{2}
\;\;,\;\;
Y_{\mu} = \frac{ B_{\mu} {\tau}_3}{2}
\label{fields}
\end{equation}
The covariant derivative $D_{\mu}U$ and the covariant field strength
tensors are defined as
\begin{eqnarray}
D_{\mu}U &=& {\partial}_{\mu} U + i g W_{\mu} U - i g' U Y_{\mu}
\nonumber
\\ F_{\mu\nu}(x) & = & {\partial}_{\mu} W_{\nu}(x) - {\partial}_{\nu}
W_{\mu}(x) + ig [W_{\mu}(x),W_{\nu}(x)]  \nonumber \\
B_{\mu\nu}(x) & =&  {\partial}_{\mu} B_{\nu}(x) - {\partial}_{\nu}
B_{\mu}(x)
\label{cov}
\end{eqnarray}
${\cal L}_{GF}$, and ${\cal L}_{FP}$ in eq.(\ref{NL})
denote the gauge fixing and Fadeev Popov Lagrangians respectively
which in the present work will be chosen in the Landau gauge
\cite{4}.  The ${\cal L}_i$ terms in eq.(\ref{EChL}) are the $SU(2)_L
\times U(1)_Y$ invariant functions of the gauge vector bosons and the GB
fields, other than ${\cal L}_{NL}$, and are made of 1 term of
dimension two, ${\cal L}_0$, and 13 terms of dimension four, ${\cal
L}_i, i=1,13$.	The electroweak chiral parameters $\alpha_i$ appear
in the definition of the ${\cal L}_i$ terms which can be written as
follows \cite{4}:
\begin{eqnarray}
{\cal L}_0  & = & \frac{1}{4} g^2 {\alpha}_0 v^2 [Tr(TV_{\mu})]^2
\nonumber   \\
{\cal L}_1  & = & \frac{1}{2} g^2 {\alpha}_1
B_{\mu\nu} Tr(TF^{\mu\nu})
\nonumber   \\
{\cal L}_2  & = & \frac{1}{2} i g {\alpha}_2 B_{\mu\nu}
Tr(T[V^{\mu},V^{\nu}]) \nonumber   \\
{\cal L}_3  & = & i g {\alpha}_3 Tr(F_{\mu\nu}[V^{\mu},V^{\nu}])
\nonumber   \\
{\cal L}_4  & = & {\alpha}_4 [Tr(V_{\mu}V_{\nu})]^2   \nonumber  \\
{\cal L}_5  & = & {\alpha}_5 [Tr(V_{\mu}V^{\mu})]^2   \nonumber  \\
{\cal L}_6  & = & {\alpha}_6 Tr[(V_{\mu}V_{\nu})]
Tr(TV^{\mu}) Tr(TV^{\nu})  \nonumber \\
{\cal L}_7  & = & {\alpha}_7 Tr[(V_{\mu}V^{\mu})] [Tr(TV^{\nu})]^2
\nonumber   \\
{\cal L}_8  & = & \frac{1}{4} g^2 {\alpha}_8 [Tr(TF_{\mu\nu})]^2
\nonumber   \\
{\cal L}_9  & = & \frac{1}{2} i g {\alpha}_9 Tr(TF_{\mu\nu})
Tr(T[V^{\mu},V^{\nu}]) \nonumber   \\
{\cal L}_{10}  & = & \frac{1}{2} {\alpha}_{10}
[Tr(TV_{\mu})Tr(TV_{\nu})]^2  \nonumber  \\
{\cal L}_{11} & = & {\alpha}_{11} Tr[({\cal D}_{\mu} V^{\mu})^2]
\nonumber  \\
{\cal L}_{12} & = & \frac{1}{2} {\alpha}_{12} Tr(T{\cal D}_{\mu}
{\cal D}_{\nu}V^{\nu}) Tr(TV^{\mu})  \nonumber	\\ {\cal L}_{13} & =
& \frac{1}{2} {\alpha}_{13} [Tr(T{\cal D}_{\mu} V_{\nu})]^2
\label{Li}
\end{eqnarray}
where
\begin{eqnarray}
T& =& U {\tau}^3 U^{\dagger}\;\;,\;\; V^{\mu}\; =\; (D^{\mu}U)
U^{\dagger}
\nonumber  \\
{\cal D}_{\mu} O(x)&  = & {\partial}_{\mu} O(x) + ig [W_{\mu}(x), O(x)]
\end{eqnarray}

\section{Searching for non-resonant strongly interacting
$VV$ signals}

In the following we will analyze the possibilities of measuring the
EChL parameters $\alpha_i$ at LHC, by means of
$W^{\pm}Z$ and $ZZ$ production, with the $W's$ and $Z's$ decaying
into the cleanest leptonic channels: $W\rightarrow
\nu_ee,\nu_{\mu}\mu$ and $Z\rightarrow e^+e^-,\mu^+\mu^-$.  It
implies a reduction on the $VV$ number of events given by the
leptonic branching ratios: $BR(WZ)=0.013$ and $BR(ZZ)=0.0044$
respectively.  All the number of events reported here include these
reduction factors.

The parameters for LHC have been chosen as follows: The $pp$ center
of mass enegy is $\sqrt{s}=14TeV$ and the luminosity is ${\cal
L}=10^{34}cm^{-2}s^{-1}$. The number of events presented in this work
correspond to an integrated luminosity of $L=3\times 10^5 pb^{-1}$.

The hadronic decays of the gauge bosons have not been considered here
but a careful study including severe cuts on the final jets could
provide additional valuable information. We have postponed also for
later studies the case of like-sign $W^{\pm}W^{\pm}$ production
\cite{18}. The most problematic channel is $W^+W^-$ due to
the overwhelming background from top-antitop production with the top
quarks decaying into $W's$ and will not be studied here.

We consider the following subprocesses contributing to $W^{\pm}Z$ and
$ZZ$ production respectively. All but the last one are considered
here at tree level:
\begin{itemize}
\item[(1)]
$q \bar{q}'\rightarrow W^{\pm}Z$
\item[(2)]
$q \bar{q}\rightarrow ZZ$
\item[(3)]
$W^{\pm}Z\rightarrow W^{\pm}Z$
\item[(4)]
$W^{\pm}\gamma \rightarrow W^{\pm}Z$
\item[(5)]
$W^+W^-\rightarrow ZZ$
\item[(6)]
$ZZ \rightarrow ZZ$
\item[(7)]
$gg \rightarrow ZZ$
\end{itemize}

The complete set of helicity amplitudes for the processes (1) to (6),
corresponding to all possible helicity states of the initial and
final electroweak gauge bosons, have been computed analitically
using the  EChL in \cite{17}. These amplitudes (except
incidentally (2)) are functions of the chiral parameters $\alpha_i$.
In addition, we have considered the subprocess number (7) which is
known to give a non-negligible contribution to $ZZ$ production
\cite{19}. It takes place in the SM {\it via} one-loop of
quarks.  For numerical computations the mass of the top quark in the
loop has been fixed to $m_t=170 GeV$.

The quark-antiquark annihilation processes, (1) and (2), and the
so-called  $VV$ fusion processes ($V=W^{\pm},Z$), (3) to (6), have a
very different final-state kinematics. In the later the expectator
quark jets are left behind when the incoming quarks radiate the
initial $V$'s that then scatter. Thus, one could presumably separate
the two kind of processes by requiring a tagged forward jet.  In
fact, a big effort is being done by the experimental physicists
comunity in this concern.  As we will see later on, the forward jet
tagging  may play an important role in searching for strongly
interacting $VV$ signals since they  mainly (but not
only) manifest in $VV$ fusion processes. For comparison, we will present
here our results for the two possibilities,  both without
and with jet tagging respectively.

A FORTRAN code has been written \cite{17} that implements all the
helicity amplitudes for the above (1) to (7) subprocesses and adds the
appropriate combinations to provide a numerical prediction for
producing all the possible final polarization states:
$V_LV_L$,
$V_LV_T$,$V_TV_L$,$V_TV_T$.
We believe that it may be interesting to
study them separately in case there is some possibility of
discriminating
a longitudinal from a transverse $V$ experimentally. However, in this
work we have not profited from this possibility  and the final
polarization states have been added to provide a total number of $VV$
unpolarized events. On the other hand, the contributions from the
various initial polarization states,
$V_LV_L$,
$V_LV_T$,$V_TV_L$ and $V_TV_T$, in the fusion processes,
must be computed separately since the
structure functions for $V_L$ and $V_T$ are different.

In order to connect the subprocesses above to the $pp$ initial state
we have used the effective $V$ approximation \cite{20}, the
Weizsaker-Williams approximation \cite{21} in the case of the $\gamma$
iniciated subprocess (4), and the EHLQ (set II) structure functions
\cite{22}. Other more realiable structure functions, as the  MRSD-
\cite{23} and the GRVHO \cite{24}, have also been considered in the
literature \cite{17}, but we do not expect the results on the
accesible range for the chiral parameters at LHC to be very affected
by our choice of the structure functions. The total number of
events does depend, however, on the choice of the structure functions
but,  hopefully,
by the time LHC will start working they will be known with
precision enough as to eliminate this kind of uncertainties.

The FORTRAN program gives, in summary, the total number of expected
gold-plated  events (including the important background from process
(7)) for a given set of values of the chiral parameters $\alpha_i$
and for a given set of cuts on the subprocess variables, namely, the
maximum  $VV$ invariant mass, $M_{VV}^{max}$, the minimum
transverse momentum of the final $Z$, $P_{Tmin}^Z$, and the maximun
rapidity of the final $V$, $y_{max}^V$.

In the present work we have restricted ourselves, for simplicity, to
the minimal set of parameters corresponding to the operators that
should be included to absorb the one-loop divergences of the lowest
order lagrangian \cite{4}, namely, $\alpha_0$, $\alpha_1$,
$\alpha_2$, $\alpha_3$, $\alpha_4$ and $\alpha_5$ (all together named
here $\alpha$).  In addition, we have set in this paper the following
minimal cuts
\begin{equation}
  M_{VV}^{max}=1.5TeV,\hspace{8mm}
  P_{Tmin}^Z=300GeV,\hspace{8mm} y_{max}^V=2,
\label{cuts}
\end{equation}
but the FORTRAN code is prepared to analyse the complete set of
chiral parameters as well as to produce, starting from the minimal
cuts in eq.(\ref{cuts}), a new set of optimal cuts for each
particular channel.

Finally, we have to decide what is the signal of this non-resonant
model and what is the background. Clearly, there is not a unique
definition.  We have proceeded in two ways.
\vspace{0.5cm}

 {\bf(A)} We have compared the predictions for $VV$ production from
the EChL for a given value of each $\alpha_i$ parameter with respect
to a reference model where the parameters are all set to zero
(incidentally, this model is equivalent to the SM at tree level with
a Higgs of infinite mass).  Let $N(\alpha_i)$ be the number of
gold-plated events obtained from the EChL for the given $\alpha_i\neq
0$ value (the rest of the other parameters are set to zero).

Let $N(0)$ be the
corresponding number of events for the reference model with
$\alpha=0$. In both rates we take the minimal cuts of
eq.(\ref{cuts}).

We define the statistical significance of the signal due to a
$\alpha_i\neq0$ effect by means of the following variable:
\begin{equation}
 r_i=\frac{\mid  N(\alpha_i)-N(0)  \mid}{\sqrt{N(0)}}
\label{ri}
\end{equation}
Obviously, the larger the value of $r_i$ the better the
sensitivity of LHC to this particular parameter $\alpha_i$.
\vspace{0.5cm}

{\bf(B)} We have also compared our predictions for $N(\alpha_i)$ as
defined in (A) with the corresponding SM predictions for the
gold-plated events in the case of a light Higgs boson, $M_H=100GeV$,
and in the case of a heavy Higgs boson, $M_H=1TeV$.
We have computed the SM amplitudes for the subprocesses
above (1) to (6) at tree level. The Higgs particle contributes mainly
{\it via} the fusion processes (3), (5) and (6). For numerical
computations,
we have included the Higgs width just in the Higgs-s-channel of the
processes (5) and (6).

The comparison
with the light Higgs case is interesting since in so doing we are
comparing two typical scenarios, a strongly  and a
weakly interacting one.  The comparison with the $M_H=1TeV$ case
teaches us how much our models separate from the 'standard reference
model' of a strongly interacting  heavy Higgs.

Let $N_{SM}(M_H)$ be
the SM predictions for the given $M_H$ value. The same minimal cuts
of eq.(\ref{cuts}) are also applied here.

The significance of a $\alpha_i\neq 0$ effect is defined as:
\begin{equation}
s_i=\frac{\mid N(\alpha_i)-N_{SM}(M_H) \mid}{\sqrt{N_{SM}(M_H)}}
\label{si}
\end{equation}

\section{Results and conclusions}

Our results for the number of gold-plated  events are summarized in
Tables 1 to 7 and in Figs. 1 to 4.  The predictions from the
non-resonant EChL model, $N(\alpha_i)$, for various choices of the
most relevant parameters, $\alpha_5$, $\alpha_4$ and $\alpha_3$ are
presented in Tables 3 to 7.  The various contributions to both
channels $W^{\pm}Z$ and $ZZ$ that come from the different processes
have been presented separately, for illustration.

We have also scanned the other parameters, $\alpha_0$, $\alpha_1$ and
$\alpha_2$, but it turns out that none of them (for the moderate
values of $\alpha_i\leq 10^{-2}$ being studied) give a significant
effect.

Firstly, what we can learn from the tables when looking at the values
of the variables $r_i$ and $s_i$ is that there are, indeed,
significant effects for the case of $\alpha_3$, $\alpha_4$ and
$\alpha_5$.  These effects could be more important, of course, if
more sizeable values of the parameters were scanned, but, as we have
said already, we have preferred here to take the most plausible
values from the theoretical point of view which are of the order of
or even smaller than $10^{-2}$.

Second, we also see from the tables that LHC will be more sensitive
to the parameter $\alpha_3$ through the study of the $q\bar{q'}$
annihilation processes, whereas the parameters $\alpha_4$
and $\alpha_5$ will be tested more clearly through the $VV$ fusion
processes.  A previous study on $\alpha_3$ through $q\bar{q'}$
annihilation processes was done in [25].
On the other hand, ${\alpha}_3$ can be related to the usual
parameters, ${\kappa}_V$, $V=Z,\gamma$, for anomalous $VW^+W^-$
couplings whose effects on the annihilation processes have already
been studied by other authors [26].

The sensitivity to the $\alpha_4$ and $\alpha_5$ parameters is high
for the moderate values of $\alpha_4$ or $\alpha_5$ equal to $\pm
10^{-2}$. For instance, if we compare the predicted rates for $ZZ$
production from the EChL for $\alpha_5=10^{-2}$ (the rest of the
$\alpha_i$'s are set to zero) with the predictions from the reference
model with $\alpha=0$, we get (see Table 3) an effect with a high
statistical significance given by $r_5=6.2$. The same can be said for
$\alpha_4$ if we compare, for instance, the predictions for
$W^{\pm}Z$ production from the EChL for $\alpha_4=-10^{-2}$ (the rest
of the $\alpha_i$'s are set to zero) with the reference model rates;
we get (see Table 4) $r_4=8.3$.  The improvement in the sensitivity
to these two parameters if a $100\%$ efficient jet tagging is achieved
is obvious from Tables 3 and 4 (in this case one compares just the rates
coming from genuine $VV$ fusion processes).  Thus, for the above
chosen values, $\alpha_5=10^{-2}$ and $\alpha_4=-10^{-2}$, the
statistical significance increases to $r_5=11.4$ and $r_4=13.0$
respectively.

Finally, it is also interesting to remark (see Tables 6 and 7) that
the predictions from the EChL  are clearly
different than the predictions from the SM at tree level, with	either a
light
or a heavy Higgs, and they are particularly separated in the case of
$M_H=100GeV$, as expected.

Perhaps, the comparison with the SM with a light Higgs boson is the
most interesting one since it represents a typical weakly interacting
scenario and we want to discriminate it as much as possible from the
strongly interacting possibility which we represent with the
EChL for the given value of $\alpha_i$.  As can be seen from Tables 6
and 7, the sensitivity to $\alpha_4$ and $\alpha_5$ is quite high for
the values of $\alpha_4$ or $\alpha_5$ equal to $\pm 10^{-2}$. For
instance, when comparing with the SM with $M_H=100GeV$, we get for
$\alpha_5=10^{-2}$  a statistical significance
in the $ZZ$ channel of $s_5=7.3$ that increases to $s_5=15.3$ if jet
tagging is considered.	Similarly, for $\alpha_4=-10^{-2}$ in the
$W^{\pm}Z$ channel, we get $s_4=10.5$ and $s_4=17.7$ without and with
jet tagging respectively.  The sensitivity is still reasonably high
for the smaller values	of $\alpha_4$ or $\alpha_5$ equal to $\pm
5\times 10^{-3}$. Thus, for $\alpha_5=5\times10^{-3}$ we get, in the
$ZZ$ channel, $s_5=3.0(6.3)$ without (with) jet tagging, and for
$\alpha_4=-5\times10^{-3}$ we get, in the $W^{\pm}Z$ channel,
$s_4=4.8(8.2)$ without (with) jet tagging.  

On the other hand, and in order to provide information on the
different kinematical structures of the final states for the various
processes considered here, we have also produced some plots (see
Figs.1 to 4) with the distributions of the gold-plated events in the
$M_{VV}$ and $P_T^Z$ variables. We have chosen here some particular
values of the $\alpha_i$ parameters for illustration.  We see from
the figures that choosing optimal cuts in these variables (or the
corresponding ones of the final leptons) one could improve
considerably the sensitivity to the parameters. Probably, a higher
value on $P_{Tmin}^Z$ (and/or $M_{VV}^{min}$) would help us in this
concern.

In summary, after a systematic scanning of the electroweak chiral
parameters $\alpha_0$, $\alpha_1$, $\alpha_2$, $\alpha_3$, $\alpha_4$
and $\alpha_5$ we conclude that LHC will be sensitive to three of
them, $\alpha_3$, $\alpha_4$ and $\alpha_5$ by analysing the leptonic
gold-plated events in $W^{\pm}Z$ and $ZZ$ production. Furthermore, the
sensitivity to $\alpha_4$ and $\alpha_5$ will improve considerably if
forward jet tagging is achieved.  Finally, in order to give the range
of the chiral parameters values that will be accessible at LHC we
need to fix a criterion to define whether an effect (signal)
due to a given $\alpha_i \ne 0$ is statistically significant or not.
For instance, if we define that a signal due to $\alpha_i \ne 0$ is
statistically significant whenever the variables in eqs.(\ref{ri})
and eqs.(\ref{si}) satisfy $r_i$ or $s_i\ge 5$ then we conclude
from the present study
that the following range of chiral parameters will be acessible
at LHC: $$|\alpha_3|\ge 10^{-2}\;\;,\;\; |\alpha_4|\ge
10^{-2}\;\;,\;\; |\alpha_5|\ge 10^{-2}.$$
If instead, we relax this criterion to the condition $r_i$
or $s_i \ge 3$
then the following more ambitious range will be
reached:
$$|\alpha_3|\ge 5 \times 10^{-3}\;\;,\;\;
|\alpha_4|\ge 5\times
10^{-3}\;\;,\;\; |\alpha_5|\ge 5\times 10^{-3}.$$

The acessible range of the above chiral parameters at LHC will be
enlarged in at least one order of magnitud respect to the present
acessible range at LEP [27].

\section*{Acknowledgments}
We would like to thank the TH division at CERN for financial support
and for their kind hospitality. We appreciate the continous
encouragement from D.Denegri and thank him for the various
discussions and useful comments on this and related subjects. We
thank T. Rodrigo and the CMS collaboration for discussions. This work
has been partially supported by the spanish Ministerio de Educacion y
Ciencia under projects CICYT AEN93-0673, AEN90-0034 and AEN93-0776.

\newpage

\section*{Figure Captions}

\vspace{1 cm}

\noindent Figure 1. Sensitivity to $\alpha_5$ in the $M_{VV}$ and
$P_T^Z$ distributions of gold-plated events.  We have applied the
minimal
cuts as defined in the text. The solid lines are the
predictions for the fusion processes in the non-resonant EChL model
for two different values of the $\alpha_5$ parameter (the other
parameters are set to zero).  The dashed lines are the predictions
for the most relevant background processes. They include  the rates
from $q\bar{q}'$ annihilation in the case of the $WZ$ channel, and
the rates from $q\bar{q}$ and gluon fusion processes in the case of
the $ZZ$ channel.

\bigskip

\noindent Figure 2. The same as in Figure 1, but for two different
values
of the ${\alpha}_4$ parameter (the other parameters are set to zero).

\bigskip

\noindent Figure 3. Sensitivity to $\alpha_3$ in the $M_{WZ}$ and
$P_T^Z$ distributions of gold-plated events. We have applied the minimal
cuts as defined in the text.
Only rates from the $q\bar{q}'\rightarrow W^{\pm}Z$
process, where the effect of $\alpha_3$ is larger, are shown. The
solid lines are the predictions in the non-resonant EChL model for
two values of $\alpha_3$ (the other parameters are set to zero). The
dashed lines are the predictions for the background in the SM.

\bigskip

\noindent Figure 4.  $M_{WZ}$ and $P_T^Z$ distributions of
gold-plated events in the Standard Model with a light ($M_H=100 GeV$)
and a heavy ($M_H=1TeV$) Higgs particle. We have applied the minimal
cuts as defined in the text.
The solid lines are the predictions from the fusion
processes. The dashed lines are the predictions from the $q{\bar q'}$
process.

\newpage
\begin{table}[h]
\caption{Gold-plated event rates for the reference model with
${\alpha}=0$.
}
\vspace{0.7cm}
\centering
\begin{tabular}{|l|c|} \hline
  & Reference Model    \\
  &  ($\alpha =0$)     \\   \hline
$q\bar{q}'     \rightarrow W^{\pm}Z^0$  &  197   \\   \hline
$W^{\pm}Z^0    \rightarrow W^{\pm}Z^0$ &   88  \\   \hline
$W^{\pm}\gamma \rightarrow W^{\pm}Z^0$ &   48  \\   \hline
total $W^{\pm}Z^0$   &	333  \\   \hline\hline
$q\bar{q}      \rightarrow  Z^0Z^0$  &	 40   \\   \hline
$gg    \rightarrow  Z^0Z^0$  &	 17  \\   \hline
$W^+W^-  \rightarrow  Z^0Z^0$  &   24  \\   \hline
$Z^0Z^0  \rightarrow  Z^0Z^0$  &    0  \\   \hline
total $Z^0Z^0$	  &   81  \\   \hline
\end{tabular}
\end{table}

\begin{table}[h]
\caption{Gold-plated event rates in the Standard Model with a light
and a heavy Higgs boson. The top quak mass is fixed to $m_t=170 GeV$}
\vspace{0.7cm}
\centering
\begin{tabular}{|c|c|c|} \hline
  & SM	& SM	\\
  &  ($m_H=100GeV$)  &	($m_H=1 TeV$)  \\  \hline
fusion $W^{\pm}Z^0$  & 106   & 119  \\ \hline
$q\bar{q}' \rightarrow W^{\pm}Z^0$  &  197  & 197  \\  \hline
total $W^{\pm}Z^0$   & 303   & 316  \\ \hline\hline
fusion $Z^0Z^0$  & 17 & 43  \\	\hline
$q\bar{q} \rightarrow Z^0Z^0$  &  40  & 40  \\ \hline
$gg \rightarrow Z^0Z^0$  &  17 & 17  \\  \hline
total $Z^0Z^0$	& 74 & 100  \\	\hline
\end{tabular}
\end{table}

\clearpage

\begin{table}[h]
\caption{Sensitivity to ${\alpha}_5$ (the other parameters are
set to zero). Comparison with the reference model, ${\alpha}=0$. Only
the predictions for the gold-plated events rates which are different
than in the reference model are shown explicitely. Total rates
include all the contributions from $VV$ fusion, annihilation
processes, and, for final state $Z^0Z^0$, the $gg$ fusion too. The
quantities in parenthesis are the corresponding predictions if a
$100\%$ efficient jet tagging is considered.}
\vspace{0.7cm}
\centering
\begin{tabular}{|l|c|c|c|c|c|c|}   \hline
    &  \multicolumn{6}{c|}{${\alpha}_5$} \\  \cline{2-7}
 & $10^{-2}$ & -$10^{-2}$ &  $5 \times 10^{-3}$ &  -$5 \times
10^{-3}$ & $10^{-3}$ & -$10^{-3}$ \\  \hline $W^{\pm}Z^0
\rightarrow W^{\pm}Z^0$  & 67 & 173 & 69 & 122 & 83 & 93
\\  \hline
total $W^{\pm}Z^0$ &  312 &  418 & 314 & 367 & 328  & 338   \\
\hline ${r_5 |}_{W^{\pm}Z^0}$ & 1.2 (1.8) & 4.7 (7.3) & 1.0(1.6) &
1.9 (2.9) & 0.3(0.4) & 0.3 (0.4)  \\  \hline\hline $W^+W^-
\rightarrow  Z^0Z^0$ &	62  & 21 &  39 &  18 &	26 & 22  \\
\hline
$Z^0Z^0  \rightarrow  Z^0Z^0$ &  18 & 18 & 4 & 4 & $\sim$ 0 & $\sim$
0
\\ \hline
total $Z^0Z^0$ & 137  &  96 & 100 &  79 &  83 &  79 \\	\hline ${r_5
|}_{Z^0Z^0}$  & 6.2 (11.4) & 1.7 (3.1) & 2.1 (3.9) & 0.2 (0.4) & 0.2
(0.4) & 0.2 (0.4)  \\  \hline
\end{tabular}
\end{table}
\begin{table}[h]
\caption{The same as in Table 3, but for the ${\alpha}_4$ parameter.}
\vspace{0.7cm}
\centering
\begin{tabular}{|l|c|c|c|c|c|c|}   \hline &
\multicolumn{6}{c|}{${\alpha}_4$} \\  \cline{2-7} & $10^{-2}$ &
-$10^{-2}$ &  $5 \times 10^{-3}$ &  -$5 \times 10^{-3}$ & $10^{-3}$ &
-$10^{-3}$ \\  \hline $W^{\pm}Z^0   \rightarrow W^{\pm}Z^0$ & 109 &
240 & 81 & 142 & 82 & 95
\\  \hline
total $W^{\pm}Z^0$ &  354 &  485 & 326 & 387 & 327  & 340   \\
\hline ${r_4 |}_{W^{\pm}Z^0}$ & 1.2 (1.8) & 8.3 (13.0) & 0.4(0.6) &
3.0 (4.6) & 0.3 (0.5) & 0.4 (0.6)  \\  \hline\hline $W^+W^-
\rightarrow  Z^0Z^0$ &	36  & 20 &  28 &  21 &	25 & 24  \\
\hline
$Z^0Z^0  \rightarrow  Z^0Z^0$ &  18 & 18 & 4 & 4 & $\sim$ 0 & $\sim$
0
\\    \hline
total $Z^0Z^0$ & 111  &  95 & 89 &  82 &  82 &	81   \\  \hline ${r_4
|}_{Z^0Z^0}$  & 3.3 (6.1) & 1.6 (2.9) & 0.9 (1.6) & 0.1 (0.2) & 0.1
(0.2) & $\sim$ 0  ($\sim$ 0)	\\  \hline
\end{tabular}
\end{table}

\begin{table}[h]
\caption{Sensitivity to ${\alpha}_3$. Comparison with the reference model
, ${\alpha}=0$. }
\vspace{0.7cm}
\centering
\begin{tabular}{|l|c|c|}   \hline
    &  \multicolumn{2}{c|}{${\alpha}_3$} \\  \cline{2-3} & $10^{-2}$
& -$10^{-2}$	\\  \hline $q\bar{q}' \rightarrow W^{\pm}Z^0$  &  149
& 284  \\  \hline $W^{\pm}Z^0 \rightarrow W^{\pm}Z^0$  &  90  & 86
\\ \hline $W^{\pm} \gamma \rightarrow W^{\pm}Z^0$ &  48  &  47	\\
\hline total $W^{\pm}Z^0$ &  287 &  417  \\  \hline ${r_3
|}_{W^{\pm}Z^0}$ & 2.5	& 4.6	\\  \hline\hline

$W^+W^-  \rightarrow  Z^0Z^0$ &  25  & 24    \\   \hline total
$Z^0Z^0$ & 82  &  81 \\  \hline ${r_3 |}_{Z^0Z^0}$  & 0.1   &  $\sim$
0   \\	\hline
\end{tabular}
\end{table}

\clearpage

\begin{table}[h]
\caption{Sensitivity to ${\alpha}_4$ and ${\alpha}_5$ in
$W^{\pm}Z^0$ production (the rest of the other parameters are set to
zero). Comparison with the Standard Model with a light Higgs particle
($M_H=100GeV$) and with a heavy Higgs ($M_H=1 TeV$).}
\vspace{0.7cm}
\centering
\begin{tabular}{|l|c|c|c|c|c|c|}   \hline
 $W^{\pm}Z$ & \multicolumn{2}{c|}{${\alpha}_5$} &
\multicolumn{2}{c|}{${\alpha}_4$} & \multicolumn{2}{c|}{${\alpha}_3$}
\\  \cline{2-7} & $-10^{-2}$ & $-5\times 10^{-3}$ &  $-10^{-2}$ &
$-5 \times 10^{-3}$ & $-10^{-2}$ & $10^{-2}$ \\  \hline $N(\alpha_i)$
& 418 & 367 & 485 & 387 & 417 & 287
\\  \hline $s_i(M_H=1TeV)$ & 5.7(9.3) & 2.9(4.7) & 9.5(15.5) &
4.0(6.5)& 5.7 & 1.6  \\  \hline $s_i(M_H=100GeV)$
&6.6(11.2)&3.7(6.2)&10.5(17.7)&4.8(8.2) &  6.5 & 0.9  \\ \hline
\end{tabular}
\end{table}

\begin{table}[h]
\caption{ The same as in Table 6 but for
$Z^0Z^0$ production.}
\vspace{0.7cm}
\centering
\begin{tabular}{|l|c|c|c|c|}   \hline
 $ZZ$ & \multicolumn{2}{c|}{${\alpha}_5$} &
\multicolumn{2}{c|}{${\alpha}_4$} \\  \cline{2-5} & $10^{-2}$ &
$5\times 10^{-3}$ &  $10^{-2}$ &  $5 \times 10^{-3}$ \\ \hline
$N(\alpha_i)$  & 137 & 100 & 111 & 89
\\  \hline
$s_i(M_H=1TeV)$  & 3.7(5.6) & $\sim$0($\sim$0) & 1.1(1.7)  & 1.1(1.7)
\\  \hline $s_i(M_H=100GeV)$ & 7.3(15.3) & 3.0(6.3) & 4.3(9.0) &
1.7(3.6) \\ \hline
\end{tabular}
\end{table}

\clearpage

\begin{figure}[h]
\let\picnaturalsize=N
\def\picsize{6in}
\def\picfilename{rota5.ps}
\ifx\nopictures Y\else{\ifx\epsfloaded Y\else\input epsf \fi
\let\epsfloaded=Y
\centerline{\ifx\picnaturalsize N\epsfxsize \picsize\fi
\epsfbox{\picfilename}}}\fi
\label{fig1}
\end{figure}


\clearpage

\begin{figure}[h]
\let\picnaturalsize=N
\def\picsize{6in}
\def\picfilename{rota4.ps}
\ifx\nopictures Y\else{\ifx\epsfloaded Y\else\input epsf \fi
\let\epsfloaded=Y
\centerline{\ifx\picnaturalsize N\epsfxsize \picsize\fi
\epsfbox{\picfilename}}}\fi
\label{fig2}
\end{figure}


\clearpage

\begin{figure}[h]
\let\picnaturalsize=N
\def\picsize{6in}
\def\picfilename{rota3.ps}
\ifx\nopictures Y\else{\ifx\epsfloaded Y\else\input epsf \fi
\let\epsfloaded=Y
\centerline{\ifx\picnaturalsize N\epsfxsize \picsize\fi
\epsfbox{\picfilename}}}\fi
\label{fig3}
\end{figure}


\clearpage

\begin{figure}[h]
\let\picnaturalsize=N
\def\picsize{6in}
\def\picfilename{rotsm.ps}
\ifx\nopictures Y\else{\ifx\epsfloaded Y\else\input epsf \fi
\let\epsfloaded=Y
\centerline{\ifx\picnaturalsize N\epsfxsize \picsize\fi
\epsfbox{\picfilename}}}\fi
\label{fig4}
\end{figure}

\end{document}